\documentclass[aps,prl,twocolumn,superscriptaddress,floatfix]{revtex4}
\usepackage{graphicx}
\usepackage[dvips,unicode,colorlinks,linkcolor=blue,citecolor=blue,urlcolor=blue]{hyperref}
\begin{document}
\title{The simplest model of polymer crystal exhibiting polymorphism.}

\author{E. A. Zubova}

\affiliation{Semenov Institute of Chemical Physics, Russian Academy
of Sciences, Moscow 117977, Russia}
\affiliation{
Correspondence and requests for materials should be addressed to E.A.Z. (zubova@center.chph.ras.ru).
}

\author{N. K. Balabaev}
\affiliation{Institute of Mathematical Problems of Biology, Russian
Academy of Sciences, Pushchino, Moscow Oblast 142290, Russia}

\author{A. V. Savin}

\affiliation{Semenov Institute of Chemical Physics, Russian Academy
of Sciences, Moscow 117977, Russia}

\begin{abstract}
Almost all the polymer crystals have several polymorphic modifications. Their
structure and existence conditions, as well as transitions between them are not understood even
in the case of the 'model' polymer polyethylene (PE). For analysis of polymorphism in polymer
crystals, we consider the simplest possible model of polymer chain: an extended flat zigzag
made of 'united' atoms (replacing CH$_2$-groups in PE chain); the united atoms belonging to
different zigzags interact via Lennard-Jones potential. Analysis of potential of interaction
between such zigzags allowed to predict the structure of five possible equilibrium lattices in
polymer crystal built out of such zigzags. Molecular dynamics simulation of the crystal built
out of flexible zigzags showed that, depending on model parameters (dimensions of the zigzag
and equilibrium distance of Lennard-Jones potential), one to three of these lattices are stable
in bulk at low temperatures. We have determined the model parameters at which the existing
stable lattices are analogous to the ones observed in real PE and linear alkanes. The triclinic
lattice has the lowest potential energy, then follows the monoclinic lattice, and the
orthorhombic lattice has the highest energy, exactly as in full atomic (with hydrogens)
molecular dynamics models.
\end{abstract}
\maketitle

\section{Introduction}
\label{intro}
Polymorphism is an attribute of polymer crystals. It is pronounced even in the
simplest, 'model' polymer polyethylene (PE) [-CH$_2$-]$_n$. The stable low-temperature phase in
its linear odd n-alkanes is the orthorhombic (O) one, in its even n-alkanes - the triclinic (T)
one up to n=24 and the monoclinic (M) one when $26<n<34$ \cite{ancient-different-alkanes-1}. PE
melt and solution crystallization gives O folded-chain crystallites, while on substrates
\cite{PE-on-paper,PE-epitaxial-crystallization} and after polymerization inside nanochannels
\cite{PE-polymerization-in-nanochannels} there is a large amount of non-O (M and T)
crystallites. At higher temperatures, one can also see a series of rotator phases in n-alkanes
\cite{rotator-phases-alkanes}, and the hexagonal conformationally disordered phase in PE
\cite{condis}.

There exists the long history of study of PE phases depending on pressure and temperature.
Starting with the works by Seto \cite{Seto-mono-under-pressure-1968}, the transitions between
orthorhombic and monoclinic phases were widely investigated (see, for example,
\cite{Takahashi-temperature-1988,Lotz-mono-crystallization-1989,Aulov-2004} and references in
them; for a review, see \cite{quasi-review-1997}). On the other hand, the P-T phase diagram
near the triple point melt-orthorhombic-hexagonal phases of PE was very early obtained
\cite{Bassett-hexagonal-1974} and then many times specified depending on length of PE chains,
samples morphology and thermal history. But, in spite of all the efforts, the exact place of M
and T phases on this diagram has never been established; the character of their instability is
still unclear. Almost at any pressure, they normally coexist with the O phase. The pressure, at
which they appear and the rate at which they transform back to the O phase, depend on mode of
preparation of the samples (compare, for example, \cite{Seto-mono-under-pressure-1968} and
\cite{pressure-M-T-in-PE-2007}). The understanding of the structure of these lattices can shed
light on this unsolved problem.

One usually treats polymorphic modifications in molecular crystals in the framework of the
'closest packing' approach \cite{Kitaigorodskii}, considering number of 'contacts' between
atoms in different lattices. This approach, as well as molecular mechanics studies
\cite{PE-molecular-mechanics}, does not allow to say which interactions between molecules in a
lattice are principal and what balance of forces forms every lattice. Correspondingly, one can
not say which influence on the lattice is able to destroy its system of couplings and to
provide the transition to another lattice. For example, in PE, T and M phases appear under
pressure \cite{Seto-mono-under-pressure-1968,pressure-M-T-in-PE-2007}, tension
\cite{drawing-PE-single crystals} and shear deformation\cite{shear-Seguela-1998}. But the
needed directions of influence and resulting modes of deformation are still not understood.

To analyze a coupling between two molecules in a molecular crystal, one is to have their
potential of interaction in treatable (better analytical) form. In the case of PE, it is
natural to suppose that the details of the potential determined by hydrogen atoms are not
principal for the system of couplings in the lattice. Then one can replace CH$_2$-groups by
'united' atoms (UA), and the chains in the lattice - by flat extended zigzags made of the UA.

However, it is a common opinion probably since 1980th \cite{against-UA-1-Ryckaert-1985} that
'the UA approximation does not provide a good representation of the crystal structures of
alkanes' \cite{against-UA-2-Hanna-2005}. Indeed, we saw \cite{vacancies,UA-PE-hexagon} that, at
some model parameters, the only crystal phase at low temperatures is a M lattice, different
from the M lattice of PE. The other model parameters can even give a mobile hexagonal phase at
300K \cite{against-UA-2-Hanna-2005}. Therefore, all the works on crystalline phases in PE and
alkanes use the full atomic (FA) model (with hydrogens) in spite of 'an order of magnitude
increase in the processor time compared to the UA case' \cite{against-UA-2-Hanna-2005}.

The goal of this article is to analyze equilibrium crystal lattices present in UA PE model at
different model parameters, compare them with the lattices present in a realistic FA model of
PE (taking into account hydrogen atoms) and to find the parameters at which the stable in bulk
at low temperatures lattices are analogous to the ones observed in the FA model - and in real
PE and linear alkanes.

The FA model that we will use as a standard was introduced in \cite{diel-rel}. It uses
generally accepted Amber99 force field \cite{Amber} and gives the results on energies and
densities of PE lattices similar to the ones obtained in other realistic empirical force fields
\cite{PE-molecular-mechanics,Kobayashi-bad-triclinic-1979}.
\section{Geometric parameters of UA PE model.}
\label{UA-PE}
UA PE model is widely used in modelling of properties of liquid and gaseous n-alkanes. The
review \cite{UA-PE-models-review} compares different sets of constants for simulation of fluid
properties. We shall dwell on geometric parameters responsible for the presence of polymorphism
in crystalline phase.

Van-der-Waals interaction of two UA belonging to different zigzags and placed $r$ apart is
normally described by 6-12 Lennard-Jones potential:
\begin{equation}
\label{LJ}
U_{LJ} = \varepsilon \left( {\frac{{{R_0}^{12}}}{{{r^{12}}}} -
\frac{{2{R_0}^6}}{{{r^6}}}} \right)
\end{equation}
The first evident geometric parameter of the model is the degree of 'collectivity' of
interactions between UA, the ratio $r_0=R_0/c$, where $c$ is longitudinal period of the PE
chain. The more is $r_0$, the more UA of the neighboring zigzags take part in generation of the
potential near an atom \cite{F-K-model}. For example, in the Amber99 force field \cite{Amber},
for hydrogen-hydrogen, hydrogen-carbon and carbon-carbon interactions, the values $r_0$ equal
to 1.166, 1.331 and 1.497 correspondingly. This parameter is widely varied in different
simulations.

\begin{figure}[tb]
\resizebox{0.3\textwidth}{!}{%
  \includegraphics[angle=0, width=1\linewidth]{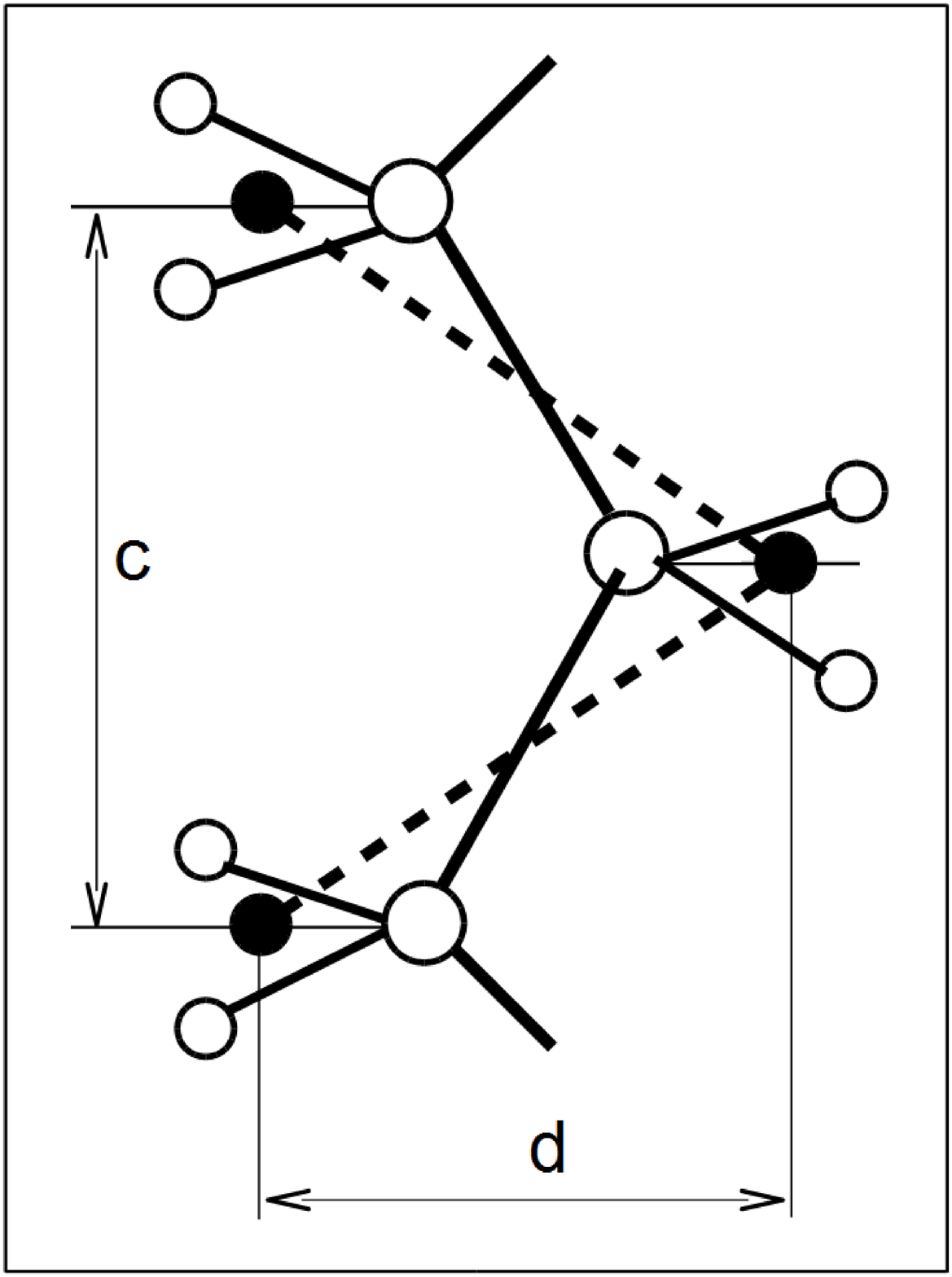}}
\caption{Choice of the UA' positions relative to the PE chain. Carbon (large) and hydrogen
(small) atoms are white, UA are black, the bonds between the UA are dashed.}
\label{fig-geometry-UA-PE-model}
\end{figure}

The second geometric parameter is the same in all but one of the works. Namely, UA are as a
rule placed on carbon atoms, although, intuitively, it seems more adequate to shift united
force centers closer to the hydrogens in the CH$_2$-groups. It has been done in model
\cite{Toxvaerd} where mass and force centers were divided (mass centers stayed on carbon
atoms). According to work \cite{UA-PE-models-review}, this model gives the best agreement with
experimental data in modelling of fluid properties of n-alkanes. We did not divide force and
mass centers of UA, and placed a united atom on the bisector of the angle between hydrogen
atoms, keeping the longitudinal period $c$ of the PE chain and changing the distance $d$
between the two rows of UA (fig. \ref{fig-geometry-UA-PE-model}). In such an approach, we
explicitly introduce the natural geometric characteristic of a flat zigzag, the ratio
$d_0=d/c$, as the second geometric parameter of the model. When a united atom shifts from
carbon to the point between hydrogens, $d_0$ changes from $0.337$ to $0.843$.

\section{Extrema of potential of interaction between two extended parallel zigzags of UA.}
\label{2zigzags}
\begin{figure}[tb]
\includegraphics [angle=0, width=1\linewidth]{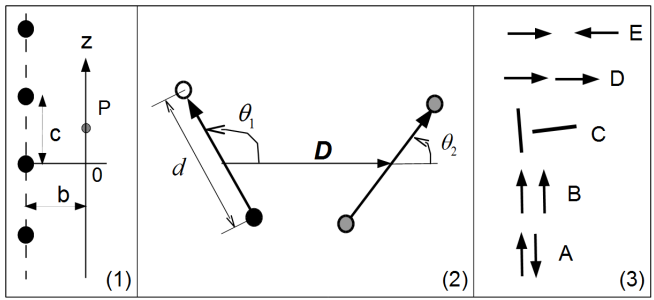}
\caption{Steps of analysis of potential of interaction between two infinite parallel extended
zigzags made of UA. (1) Finding potential $W_1$ of an infinite row of UA at the point $P$ as a
function of $b$ and $z$. (2) Finding potential of interaction $W_4$ between two zigzags. We
show the projection of the zigzags on a plane orthogonal to their axes. The arrow points from
the row of even atoms to the row of odd atoms of one zigzag. Black circle denotes the row in
which zeroth atom belongs to the projection plane, white circle - the row in which the nearest
to the projection plane atoms are at the distance $c/2$ from it. We find $W_4$ as a function of
the distance $D$ between the zigzags, their setting angles $\theta_1$ and $\theta_2$, and the
shift $z$ of the second zigzag along its axis from the projection plane. When $z=0$, the rows
of even and odd atoms of the second zigzag also become 'black' and 'white' correspondingly. (3)
Extrema of $W_4$. In the sketches, we show only the arrows introduced on step (2); here, we
imply that the rows of atoms are 'black' and 'white'. The lowest in energy coupling (the
absolute minimum) has notation A, the higher extrema - notation B, C,... and so on.}
\label{extrema}
\end{figure}

If we accept that two UA interact according to (\ref{LJ}), then, if $r_0$ is reasonably large,
$r_0>1.115$, the potential of interaction of one united atom with an infinite row of UA can be
obtained by Poisson summation formula (see fig. \ref{extrema}(1)) where the first two terms
quite suffice \cite{F-K-model}:
\begin{equation}
\label{row} W_1\left( {{k_0},\beta ,\zeta } \right) = \frac{3}{{16}} {k_0} \varepsilon \left(
\begin{array}{l}
 {J_0}\left( {{k_0},\beta } \right) + {J_1}\left( {{k_0},\beta } \right)\cos \left\langle \zeta  \right\rangle  +  \\
  + {J_2}\left( {{k_0},\beta } \right)\cos \left\langle {2\zeta } \right\rangle  + ... +  \\
  + {J_m}\left( {{k_0},\beta } \right)\cos \left\langle {m\zeta } \right\rangle  + ... \\
 \end{array} \right)
\end{equation}
where $k_0=2\pi r_0$, $\beta = 2\pi b/c$, $\quad\zeta=2\pi z/c$,
$$
J_0\left( {{k_0},\beta } \right) = \left( {{{\left( {\frac{{{k_0}}}{\beta }}
\right)}^{11}}\frac{{21}}{{32}} - 2{{\left( {\frac{{{k_0}}}{\beta }} \right)}^5}} \right).
$$
The function $J_0\left( {{k_0},\beta } \right)$ has a minimum equal to $-1.43132...$ at the
point $\beta=\beta_0=B_0 k_0$, $B_0=0.9471316...$, and the function $J_1\left( {{k_0},\beta }
\right)$ can be very good approximated by an exponent near the point $\beta_0$:
$$
J_1\left( {{k_0},\beta } \right) \approx {A_1}\left( {{k_0}} \right)\exp \left( { - \Lambda
({k_0})\left( {\beta  - {\beta _0}} \right)} \right),
$$
$$
A_1\left( {{k_0}} \right) = \frac{2}{3}{k_0}^5\exp \left( { - {\beta _0}} \right){f_1}\left(
{\beta_0,{k_0}} \right),
$$
$$
\Lambda ({k_0}) = 1 - {{{{\left. {\frac{{\partial {f_1}\left( {\beta ,{k_0}}
\right)}}{{\partial \beta }}} \right|}_{\beta  = {\beta _0}}}} \mathord{\left/
 {\vphantom {{{{\left. {\frac{{\partial {f_1}\left( {\beta ,{k_0}} \right)}}
 {{\partial \beta }}} \right|}_{\beta  = {\beta _0}}}} {{f_1}\left( {{\beta _0},{k_0}} \right)}}} \right.
 \kern-\nulldelimiterspace} {{f_1}\left( {{\beta _0},{k_0}} \right)}},
$$
$$
\begin{array}{l}
 {f_1}\left( {\beta ,{k_0}} \right) =  \\
  = {k_0}^6\frac{1}{{480{\beta ^{11}}}}\left( \begin{array}{l}
 945 + 945\beta  + 420{\beta ^2} +  \\
  + 105{\beta ^3} + 15{\beta ^4} + {\beta ^5} \\
 \end{array} \right) -  \\
 \quad  - 2\left( {{\beta ^2} + 3\beta  + 3} \right)/{\beta ^5} \\
 \end{array}
$$

To obtain the energy of interaction $W_4$ between two zigzags (per 1 united atom), one is (see
fig. \ref{extrema}(2)) to sum four terms of the type $W_1$ (\ref{row}), namely, the energy of
two successive UA of one zigzag in the presence of two rows of another zigzag (and to divide by
four). Unfortunately, only in the case of proximity of the zigzags to round cylinders $d/D \sim
d_0/r_0<<1$ the function $W_4$ can be adequately presented by the first terms of its Fourier
series in angles $\theta _1$ and $\theta _2$. This condition does not hold in models exhibiting
polymorphism in crystal lattices. Therefore, we found the extrema numerically as a sum of the
mentioned four terms.

It is evident that $\zeta=0$ or $\pi$ for equilibrium configurations. Taking $\zeta=0$ for
definiteness sake, we obtained four (physically different) stationary points (see fig.
\ref{extrema}(3)): A ($\theta _1=\pi/2$, $\theta _2=-\pi/2$), B ($\theta _1=\pi/2$, $\theta
_2=\pi/2$), D ($\theta _1=0$, $\theta _2=0$), E ($\theta _1=0$, $\theta _2=\pi$) similar to the
ones in dipole-dipole interaction between electrical or magnetic dipoles. There are four more
couplings designated as C in fig. \ref{extrema}(3). We did not draw arrows here, because all
four combinations are possible, although the setting angles are a little different.

Among the listed couplings, A is a stable minimum, the rest stationary points are saddles. B is
unstable in $z$ direction (slides into A), C - in $\theta _2$ direction (slides into B or A), D
- in any direction in ($\theta _1, \theta _2$) plane, as well as E, which is also unstable in
$z$ direction. The equilibrium distances $y_S=D_S/R_0$ and energies $w_S=W_4/E$
($E\left({r_0},\varepsilon \right) =1.6862 r_0\varepsilon$) in these configurations are listed
in Table \ref{table-energies-of-couplings}. The unit of energy $E\left({r_0},\varepsilon
\right)$ is chosen so that the first term in (\ref{row}) has minimum equal to (-1).
\begin{table}[t]
\caption{Distances $y_S=D_S/R_0$ and energies $w_S=W_4/E$ of equilibrium configurations of two
zigzags (extrema of $W_4$) listed in fig.\ref{extrema}(3)). The presented numerical values are
calculated for model parameters $G_4$ ($d_0=0.647$, $r_0=1.541$). In other cases, the extrema
and their order are the same, the numerical values slightly differ.}
\label{table-energies-of-couplings}
\begin{tabular}{ccc}
\hline\noalign{\smallskip}
extremum      &   $y_S$   &     $w_S$ \\
\noalign{\smallskip}\hline\noalign{\smallskip}
E             & ~~~1.351~~~~~     & -1.604    \\
D             & 1.323     & -1.745    \\
C             & 1.117     & -2.599    \\
B             & 0.931     & -3.681    \\
A             & 0.912     & -3.894    \\
\noalign{\smallskip}\hline
\end{tabular}
\end{table}
\section{Formation of equilibrium lattices.}
\label{lattices}
\begin{figure}[tb]
\includegraphics [angle=0, width=1\linewidth]{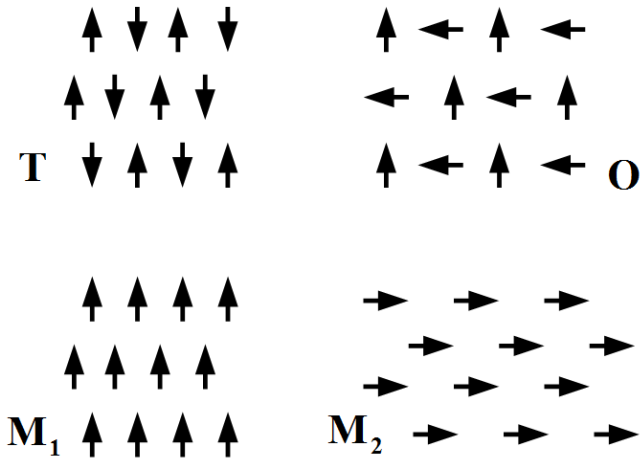}
\caption{Sketches of
possible types of equilibrium lattices built out of zigzags interacting through potential
having extrema presented in fig. \ref{extrema} (3). Every lattice consists of (horizontal) rows
of zigzags connected by equilibrium  couplings: T - A, M$_1$ - B, O - C, M$_2$ - D. In real
phases (see fig. \ref{PE-phases}), not only the couplings between rows but also the couplings
forming rows may be slightly out of equilibria.} \label{fig-lattices}
\end{figure}

We accept that structure of lattices is basically determined by interactions of a zigzag with 6
(or may be 8) zigzags of its first coordination sphere. Let us sketch a lattice as a set of
arrows (introduced in fig. \ref{extrema}) representing zigzags in the projection plane
orthogonal to the axes of the zigzags (see fig. \ref{fig-lattices}).
\subsection{Lattices based on A and B couplings: T and M$_1$.}
Geometrically, it is impossible to build a lattice using only the strongest and stable A
couplings. One can form a row of arrows (for example, horizontal) so that every two neighboring
arrows point in opposite directions (one - up, the second - down), and this will be a stable
structure; but one still is to build a two-dimensional pattern from these rows. Combining the
rows, one can not avoid forming couplings of D and E types between molecules in neighboring
rows. As a result, we get a T lattice.

In it, A couplings in horizontal rows store about twice energy than the couplings between the
rows (compare the energies of A, D and E couplings in Table \ref{table-energies-of-couplings})
and prevent the zigzags from rotation. E couplings are unstable regarding the shift of the
zigzags along their axes, and this instability is compensated by the others couplings.

T lattice built out of rows with A couplings and having couplings of D and E types between rows
may happen to be energetically less favorable than lattice M$_1$ built out of rows with B
couplings but having only D couplings between the rows.

A separate row with B couplings is an equilibrium structure, but it is unstable regarding the
shift of any zigzag along its axis. In lattice M$_1$, the rows with B couplings stabilize by
cohesion with each other: the arising couplings of D type are stable in $z$ direction and
compensate the instability of B couplings.

In both lattices T and M$_1$, the couplings D and E between rows are off their equilibria
(although the forces are small), and every zigzag needs two couplings with the neighboring row
to keep its setting angle.
\subsection{Lattices using C and D couplings: O and M$_2$.}
\begin{figure}[tb]
\includegraphics [angle=0, width=1\linewidth]{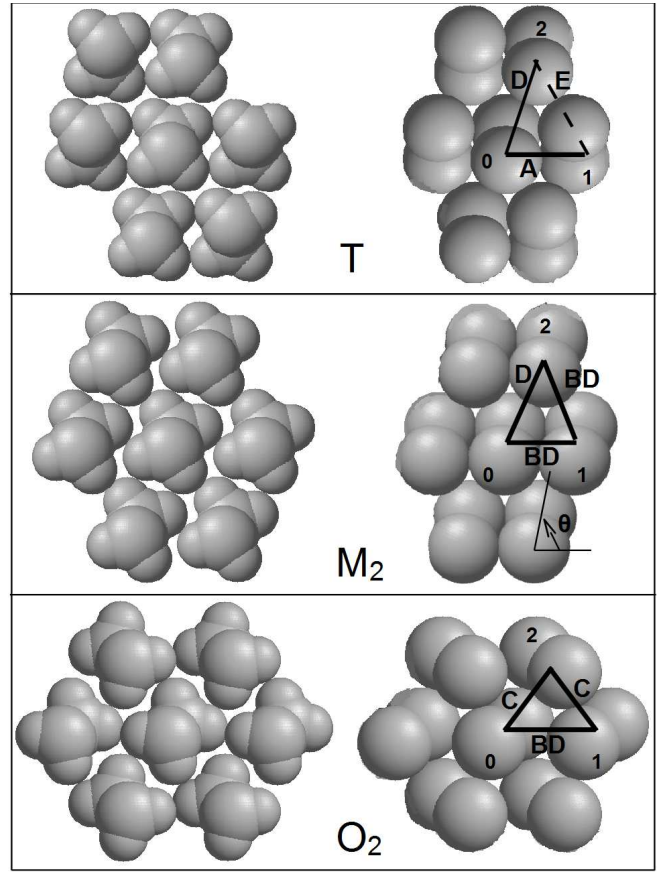}
\caption{Lattices T, M$_2$ and O$_2$ (viewed in projection on a plane orthogonal to chains'
axes) in UA PE model ($G_4$: $d_0=0.647$, $r_0=1.541$) in comparison with analogous phases of
crystalline PE (FA MD simulation, the details of the model are in \cite{diel-rel}). Radii of
the balls are equal to van-der-Waals radii of the corresponding atoms (carbon, hydrogen and
'united' ones).} \label{PE-phases}
\end{figure}

In lattice M$_2$, horizontal D couplings in rows do not form this lattice but rather are its
vulnerability. The strongest couplings are the non-equilibrium ones between rows, intermediate
between B and D, and designated as BD in fig. \ref{PE-phases}. They appear in both lattices
M$_2$ and O. Again, to maintain the equilibrium, every zigzag needs two different BD couplings
with another row. In lattice M$_2$, the setting angles of zigzags in D couplings may be off
their equilibria, and BD couplings 0-1 and 1-2 may have different lengths, and so one of these
couplings may be closer to B than another.

There can be at least two different equilibrium O lattices. In both of them, the couplings 0-2
and 1-2 (fig. \ref{PE-phases}, at the bottom) are equivalent and therefore the sum of the
setting angles of zigzags $\theta_1$ and $\theta_2$ is $\pi$. In lattice O$_1$,
$\theta_2-\theta_1<\pi/2$, in lattice O$_2$, $\theta_2-\theta_1>\pi/2$.

As we remember, the coupling C is unstable regarding rotation of zigzag 2 (see
fig.\ref{extrema}) and stable regarding rotation of zigzag 1. In accord with it, in O lattice,
the zigzag is prevented from rotation basically by its neighbors for which arrows are aligned
along the direction to the zigzag. Two other C couplings give almost flat potential, and BD
couplings only shift a little the minimum of the energy.
\section{Choice of model parameters for PE.}
\label{section-choice-of-the-point}

\begin{figure}[tb]
\includegraphics [angle=0, width=1\linewidth]{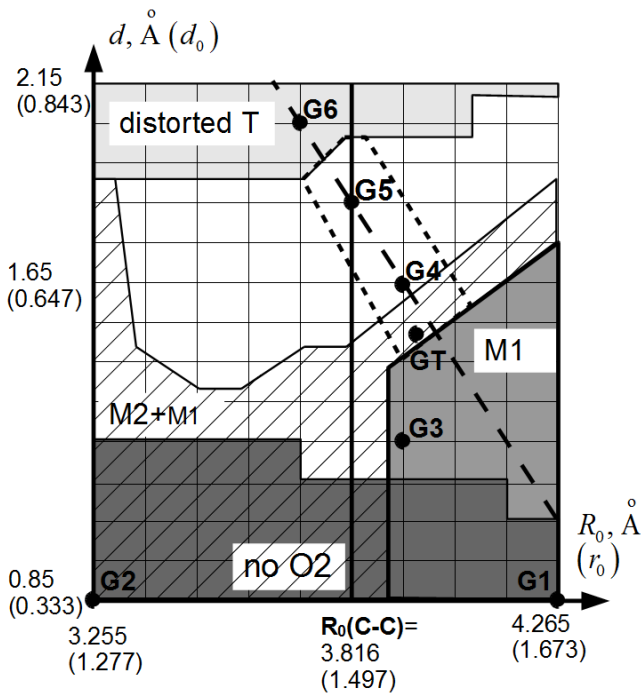}
\caption{Domains
of existence of different lattices depending on model parameters (for details, see section
\ref{section-choice-of-the-point}). The points the best corresponding to real PE lie in the
white area of the domain bounded by dashed polygon; the point G$_4$ is our best choice.}
\label{fig-choice-of-the-point}
\end{figure}

To find the point the best corresponding to PE in the rectangular area of reasonable
parameters, we selected 140 points in the rectangle (intersections of grid lines in
fig.\ref{fig-choice-of-the-point}) and studied possible equilibrium lattices in every point.

In the computational cell, we put 4 rigid zigzags parallel to $z$ axis (more exactly, two UA
from every of 4 zigzags). Every such zigzag has 4 coordinates, 3 spacial ones and the setting
angle as introduced in fig. \ref{extrema}. The first zigzag was placed so that its spacial
coordinates were zeroes, and so we got 13 variables. Two edges of the computational cell were
parallel to $x$ and $z$ axes, and so we got 3 variables characterizing the computational cell
(two lengths of the edges in ($x, y$) plane and the angle between them). The computational cell
was repeated nine times in directions of its edges in ($x, y$) plane, and 80 times in $z$
direction. In all three directions, we imposed periodic boundary conditions. For the full
Lennard-Jones energy of the obtained crystal, we solved the minimization problem in the space
of 16 variables starting at structures M$_1$, M$_2$, O$_1$, O$_2$ and T. In points G$_3$ and
G$_2$, we used the lattice parameters obtained in approximation of the first coordination
sphere with the use of analytical potential (\ref{row}). After we got the solutions in these
points, we moved to the next points using at every step as initial structures the ones obtained
in the previous points. The results are presented in fig. \ref{fig-choice-of-the-point}.

T lattice exists almost everywhere in the rectangle, excluding a couple of lowest rows of
points. In the 10\% grey domain 'distorted T', lattice T is distorted, the setting angles of
the zigzags differ from $\pi/2$ and $-\pi/2$, contrary to the experiment
\cite{Seto-mono-under-pressure-1968}, and so possible candidates for PE can not lie in the
domain 'distorted T'.

Lattice O$_2$, corresponding to O lattice of real PE, exists everywhere except for 60\% grey
domain designated as 'no O$_2$'. The setting angle $\theta_1$ changes from $\sim 30^{o}$ in the
upper part of the rectangle to $45^{o}$ at the boundary with 'no O$_2$' domain. Lattice O$_1$
is present everywhere, and has lower energy than lattice O$_2$.

Lattice M$_2$, corresponding to M lattice of real PE, exists everywhere except for the 40\%
grey domain designated as 'M$_1$' where the only minimum of M type is lattice M$_1$. In the
hatched domain 'M2+M1' these lattices coexist, and lattice M$_2$ has lower energy.

So, in the rectangle, only the points  not painted in grey can be possible candidates for
modelling of PE. Among them, the points belonging to the line $G_4$-$G_6$ have the density
equal to the density of PE at low temperatures. So, it looks reasonable to select the point for
PE in the domain bounded by a dashed polygon.

To make more precise choice, we compared the sizes of the lattices in several points on the
line $G_4$-$G_6$ and in the FA MD model. The results are presented in Table
\ref{table-G-points}.
\begin{table}[tb]
\caption{Lengths D$_{ij}$ of the couplings (in angstroms) and setting angles of the zigzags (in
degrees) in lattices T, M$_2$ and O$_2$ (fig. \ref{PE-phases}) at different model parameters
(in points shown in fig. \ref{fig-choice-of-the-point}) in comparison with FA model. In the
brackets, we point out the type of the coupling.} \label{table-G-points}
\begin{tabular}{ll|cccc|c}
\hline
\hline
  &        & G$_6$ &  G$_5$ &  G$_4$ &  G$_T$ & ~~FA PE~~  \\
\hline
T     &   D$_{01}$ (A) & ~~3.30~~  &  ~~3.40~~  &  ~~3.52~~  &  ~~3.56~~  & 3.9  \\
      &   D$_{12}$ (E) & 5.38  &  5.19  &  5.07  &  4.91  & 5.25 \\
      &   D$_{02}$ (D) & 5.13  &  5.10  &  4.99  &  4.83  & 4.37 \\
      &   $\theta_1$   & 97    &  90    &  90    &  90    & 84   \\
\hline
M$_2$ &   D$_{01}$ (B) & 3.48  &  3.56  &  3.64  &  3.64  & 4.15  \\
      &   D$_{12}$ (BD)& 4.84  &  4.81  &  4.79  &  4.70  & 4.74 \\
      &   D$_{02}$ (D) & 5.32  &  5.19  &  5.06  &  4.84  & 4.36 \\
      &   $\theta_1$   & 70    &  73    &  78    &  81    & 70   \\
\hline
O$_2$ &   D$_{01}$ (BD)& 4.80  &  4.75  &  4.68  &  4.55  & 4.82  \\
      &   D$_{12}$  (C)& 4.18  &  4.22  &  4.25  &  4.22  & 4.25 \\
      &   $\theta_1$   & 33    &  34    &  34    &  36    & 47   \\
\hline
\hline
\end{tabular}
\end{table}
One can see that the proportions of the lattices in toto approach the ones in FA model when we
move from the point G$_6$ to the point G$_T$. Besides, the dynamical behavior of the crystal
will probably be closer to the real one in the points with less $d_0$ because mass centers of
UA will be closer to carbons as it is in real PE.

The chosen dashed polygon in fig. \ref{fig-choice-of-the-point} is divided into two principally
different zones. On the left, the models have four different lattices: T (ground state), M$_2$,
O$_1$ and O$_2$; and O$_1$ was never found in real PE or alkanes. On the right, there is even
five lattices: T (ground state again), M$_2$, O$_1$, M$_1$ and O$_2$, where two 'superfluous'
lattices O$_1$ and M$_1$ are again lower in energy than the widespread at high temperatures
lattice O$_2$. The natural question is: are the found lattices really present as phases in the
model or are they somehow unstable? Indeed, our minimization problem was solved only in
16-dimensional space, and we definitely could block some modes of transformation between
lattices, and some of the lattices may prove to be saddles, and not minima.
\section{Stability of equilibrium lattices in UA PE model - MD simulation.}
\label{section-MD}
The quickest way to answer the question put in the previous section is to carry out a MD
simulation of all these lattices in large enough sample, in the framework of UA PE model having
corresponding geometrical parameters, but with normal flexible zigzags, having all mobile UA.
We took as a computational cell a parallelepiped having 120 zigzags 80 UA long packed into
expected lattices. Because we are interested in stability in bulk, we imposed periodic boundary
conditions in all three directions (and so the zigzags became 'infinite'). But, because we are
interested in the easiest conversions between lattices, we opened a way for transformations
closed in real experiments. Namely, we allowed the computational cell to change its shape (3
lengths and 3 angles) through uniform change of distances between atoms - according to the
arising stresses (local stresses averaged over the sample). So, we facilitated conversions
between lattices close in sizes - if the sample is able to preserve momentum and angular
momentum during the conversion. In real experiments, martensitic transitions in bulk normally
go through splitting samples into domains because this way of transformation is closed.

The exact values of force constants one uses in the model (the value $\varepsilon$ in
Lennard-Jones potential (\ref{LJ}), rigidities of bonds between UA, of valence and torsional
angles in zigzags) are not really important when one studies behavior of lattices at low
temperatures. We carried out the MD experiments at T=1K, and used force constants close to the
ones chosen by Toxvaerd \cite{Toxvaerd}. Our force constants are listed in
\cite{UA-PE-hexagon}, except for the value $\varepsilon$, which sometimes was taken to be equal
to 0.16 kcal/mol instead of 0.12 kcal/mol. To maintain the temperature, a collisional
thermostat \cite{thermostat-1,thermostat-2} was used with the parameters $m_0$ = 1 a.m.u. and
$\lambda= 5.5 ps^{-1}$ , and a Berendsen barostat \cite{Berendsen} was used for maintaining
pressure in 1 atmosphere.

We carried out MD simulations of all obtained in the previous section lattices at model
parameters corresponding to typical points (having different sets of equilibrium lattices) in
fig. \ref{fig-choice-of-the-point}: G$_1$, G$_2$, G$_3$, G$_T$ and G$_4$.

One gets the point G$_1$ ($d_0=0.337$, $r_0=1.673$) if one puts UA on carbon atoms and then
chooses $R_0$ so that, at temperature T=300K, the density of the resultant crystal (M$_1$)
coincides with the density of O lattice of PE. In G$_1$, we expected two lattices, M$_1$
(ground state) and O$_1$ (which is present everywhere). Lattice M$_1$ appeared to be stable,
O$_1$ transformed into ground state M$_1$. The mode of transformation is very simple and
evident; the setting angle $\theta_1$ (see fig. \ref{PE-phases}) grows up to 90$^0$ and the
setting angle $\theta_2$ drops down to 90$^0$, the sample slightly changes its sizes and we get
an accurate M$_1$ lattice. This transformation can so easily take place because the rotation of
a couple of zigzags exactly conserves angular momentum. So, in the point G$_1$, there is only
one stable phase, as was already known \cite{vacancies}.

In the point $G_2$ ($d_0=0.337$, $r_0=1.273$), with UA placed on carbons but having more
'local' interactions between atoms than in the point $G_1$, we expected the lattices M$_2$ and
again $O_1$. Both of them proved to be stable, and so, in the point $G_2$, we have two stable
phases.

In the point G$_3$ ($d_0=0.493$, $r_0=1.528$), we expected M$_1$, T, O$_1$ and O$_2$. O$_1$
transforms (exactly as in the point G$_1$) into the ground state M$_1$, the rest lattices are
stable. So, in point the G$_3$, we have 3 stable phases.

In the point G$_T$ ($d_0=0.584$, $r_0=1.549$), corresponding to the choice made by Toxvaerd
\cite{Toxvaerd} for liquid alkanes, we expected all the possible lattices in the order T,
M$_2$, O$_1$, M$_1$ and O$_2$. Only two lattices, the lowest (T) and the highest (O$_2$) in
energy proved to be stable phases, the rest lattices transformed into the ground state T. We
can conclude that the domain 'M2+M1' should be excluded from the area of possible candidates
for PE.

In the point G$_4$ ($d_0=0.647$, $r_0=1.541$), we expected T, M$_2$, O$_1$ and O$_2$. Only
lattice O$_1$ transformed into T; the rest lattices are stable.

So, in the point G$_4$, we have three low-temperature phases exactly analogous to the phases
observed in real PE (fig. \ref{PE-phases}). The ground state of the system is T phase, followed
by M$_2$ and then O$_2$, exactly as in the FA PE model. The densities of the phases differ from
the ones in the FA PE model less than by 1$\%$.
\section{Conclusion.}
Potential of interaction of two flat parallel rigid (infinite) zigzags built out of UA has 5
different extrema. Connecting zigzags by couplings close to these extrema, one can build 5
different equilibrium crystal lattices, including lattices T, M$_2$ and O$_2$ observed in PE
and alkanes at low temperatures. So, the presence of hydrogens is not necessary for the
presence of polymorphism in this polymer crystal; UA PE model catches the main features of
interaction between zigzags and provides all the needed polymorphic modifications in the
crystal built out of such zigzags. However, these equilibrium lattices can be saddles as well
as local minima in potential energy of the crystal lattice depending on geometrical parameters
of the model. In different domains of parameter space, one to five of these lattices can be
local minima.

Unfortunately, the usual choice of UA positions on carbons eliminates T lattice as a minimum
and, at (also usually used) very long-range interactions between UA, also eliminates O lattice.
This is the reason why UA model has been in disrepute. But the shift of UA to hydrogens leads
to appearance of minima for both lattices T and O$_2$.

However, a lattice can be a minimum in the space of variables of possible crystal lattices
having corresponding symmetry restrictions, but to be a saddle in full space of variables,
because the symmetry restrictions may block the modes of transformation between lattices. The
simplest way to check if an equilibrium lattice is a phase (meaning stable in bulk at low
temperatures) is to carry out a MD simulation of it in a large enough sample, with all mobile
UA. We did it for several typical points in space of model parameters and found that UA model
can have one to three phases. These three phases may be (the point G$_4$) or not be (the point
G$_3$) analogous to phases in PE. This narrowed the area for possible candidates for modelling
of PE to the white part of the domain bounded by dashed polygon in fig.
\ref{fig-choice-of-the-point}. Based on analysis carried out up to now, we would say that the
best choice for PE is point G$_4$, although, after the further investigation of temperature
behavior and phase transitions, the other points in the area can prove to be more suitable.

In the framework of UA model, one can observe phase transitions in polymer crystals in
reasonable time of calculations, which makes this model (having properly chosen parameters) a
useful instrument for studying features of polymorphic modifications and martensitic
transitions in polymer crystals.
\section{Acknowledgements.}
The work was financially supported by Russian Foundation for Basic Research (award
09-03-00230-a). We carried out MD simulations in the Joint Supercomputer Center of the Russian
Academy of Sciences. We thank prof. L.I. Manevitch for helpful discussions.

\end{document}